\documentclass[twocolumn,aps,prb,showpacs]{revtex4}
\usepackage{graphicx,epsfig,bm,color}

\begin{document}

\title{Many-Body meets QM/MM: Application to indole in water solution}
\author{A. Mosca Conte (1), E. Ippoliti (2), R. Del Sole (1), P. Carloni
(2), O. Pulci (1)}
\affiliation{(1) NAST, ETSF, INFM-SMC, CNR, Universita' di Roma Tor Vergata, Via della
Ricerca Scientifica 1, Roma \\
(2) Democritos, SISSA Scuola Internazionale Superiore di Studi Avanzati, via
Beirut 2-4, I-34014 Trieste, Italy}
\date{\today}

\begin{abstract}
Spectral properties of chromophores are used to probe complex
biological processes in vitro and in vivo, yet how the environment
tunes their optical properties is far from being fully understood.
Here we present a method to calculate such properties on large scale
systems, like biologically relevant molecules in aqueous solution.
Our approach is based on many body perturbation theory combined with
quantum-mechanics/molecular-mechanics (QM/MM) approach. We show here
how to include quasi-particle and excitonic effects for the
calculation of optical absorption spectra in a QM/MM scheme. We
apply this scheme, together with the well established TDDFT
approach, to indole in water solution. Our calculations show that
the solvent induces a redshift in the main spectral peak of indole,
in quantitative agreement with the experiments and point to the
importance of performing averages over molecular dynamics
configurations for calculating optical properties.
\end{abstract}

\maketitle

\bigskip


\bigskip

Optical properties of aromatic chromophores embody a key facet of cell
biology, allowing for a precise interrogation of a variety of biochemical
events, including signaling, metabolism and aberrant processes. These range
from probing transient interactions between biomolecules (proteins and
nucleic acids), to protein dynamics and fibrillation and plaque formation in
neurodegenerative diseases. Understanding how the environment tunes such
optical properties is therefore crucial in structural genomics, yet this
information is so far mostly lacking. A powerful tool to address this issue
is given by the so-called quantum-mechanics/molecular-mechanics (QM/MM)
methods [\onlinecite{qmmm1,qmmm2,qmmm3,qmmm4,qmmm5,qmmm6,qmmm7}]. In this
approach, the aromatic moiety is treated at quantum mechanical level, whilst
the environment is described with an effective potential: the influence of
the MM (presumably very complex and very large) environment is basically
included as an external potential, and, in case the molecule is
covalently bound to MM region, by a mechanical coupling with the environment.

Most often the QM approach is solved within Density Functional
Theory (DFT) [\onlinecite{dft1,dft2}] to study ground state
properties, and time-dependent DFT (TDDFT)
[\onlinecite{tddft1,tddft2}] when excited states are involved as in
the case of the optical properties
[\onlinecite{tddftmm1,tddftmm2,tddftmm3}]. TDDFT is computationally
very efficient, yet its predictive power depends dramatically on the
system and on the functional used to reproduce the exchange and
correlation interactions.

Several approaches, including Post-Hartree-Fock ones [\onlinecite{Olivucci}]
(configuration interaction and similar methods),
have been already used to predict optical
properties of biomolecules. Quantum-many-body techniques (MBPT)
[\onlinecite{gw1,bse}], are an attractive alternative, although of
course they come with a much higher computational cost than TDDFT. Strikingly,
however, biophysical applications of one of the most widely used
scheme, the combination of the GW method [\onlinecite{gw1}] with the
Bethe-Salpeter Equation (BSE) [\onlinecite{bse}] are so far lacking.
The GW method is used for the evaluation of the single quasiparticle
energies, and the BSE to introduce excitonic effects. Keeping in
mind future biological applications, it is imperative to assess the
accuracy of a MBPT/MM approach versus the more conventional TDDFT/MM
one.

The main assumption in interfacing a QM/MM method with TDDFT or MBPT
approaches is that the optical properties of the chromophore do not
involve the MM part's electronic structure. Hence, special care has
to be devoted to the choice of the two regions.

\begin{figure}[!h]
\includegraphics[width=7.0cm]{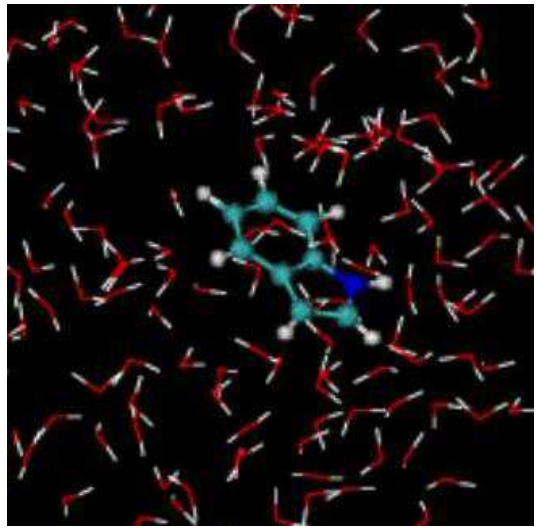} \newline
\caption{(color online) Indole in water solution. Colors
correspond to the following atomic species: BLUE=N, CYAN=C,
WHITE=H, RED=O.} \label{chdens}
\end{figure}

Here we present MBPT/MM calculations on the indole ring of the
Tryptophan protein residue (Fig. \ref{chdens}). This system appears
ideal for such an approach in several respects. First, it is very
relevant biologically, as the indole ring has been exploited as a
spectroscopic tool to monitor changes in proteins
[\onlinecite{Creed}] and to yield information about local structure
and dynamics. In fact, its spectral signatures allow it to be used
as a structural probe in proteins. Second, it contains a relatively
small number of atoms (16), which can still be treated at the GW-BSE
level.
 Next, the optical gap of liquid water (7 eV
[\onlinecite{exp-water1,exp-water2,viviana}]) is larger than the gap
of the indole molecule (4.3 eV [\onlinecite{exp_indole2}]). Under 7
eV the spectra of indole and water do not overlap, and it is
justified to treat the solvent in a classical scheme. Finally, CASPT2
calculations [\onlinecite{exp_indole2}] and
 experimental data [\onlinecite{exp_indole1}]  are available, and allow to compare the changes of
the optical properties upon passing from the gas phase to aqueous
solution.

We performed QM/MM Car-Parrinello [\onlinecite{carparrinello}]
simulations of indole in water by the fully Hamiltonian QM/MM scheme
[\onlinecite{qmmm7}]. Such scheme has been applied to a variety of
biological systems [\onlinecite{dalperaro}]. The biomolecule was
treated, at this step, at the DFT level whilst the solvent with the
Amber force field [\onlinecite{amber}]. The approach allows for an
explicit treatment of solvation, in contrast to previous studies
[\onlinecite{exp_indole1,exp_indole2,exp_indole3}].

Indole single quasiparticle energies have been then evaluated at the GW
level for several snapshots. Finally, we solved the BSE to calculate the
average absorption spectrum and compared the results with the ones obtained
within TDDFT. We calculated the indole absorbance in water as well as in gas
phase. The shift in the spectra gives the solvatochromism.

The GW approximation consists in setting the vertex in the Hedin equations
[\onlinecite{hedin}] equal to a delta function. Under this condition, the
time-Fourier transform of the proper exchange-correlation
self energy, $\Sigma(\mathbf{{r},{r^{\prime }},\omega )}$,
is a convolution of the Green function $G(\mathbf{{r},{r^{\prime }},\omega )}$,
with the screened Coulomb potential $W(\mathbf{{r},{r^{\prime }},\omega )}$.
The electronic bands are obtained by solving the following eigenproblem:

\begin{eqnarray}
\left[ -\frac{\hbar^{2} \nabla^{2}}{2m} + U^{QM}(\mathbf{r})+ U^{QM/MM}(
\mathbf{r}) + V_{H}(\mathbf{r}) \right] \phi_{j}(\mathbf{r}) &+&  \nonumber \\
\int d^{3}\mathbf{{r^{\prime}}} \Sigma(\mathbf{{r},{r^{\prime}}}
,\varepsilon^{QP}_{j}) \phi_{j}(\mathbf{r^{\prime}}) = \varepsilon^{QP}_{j}
\phi_{j}({\mathbf{r}}).&&  \label{dyson}
\end{eqnarray}

This expression is derived from the Dyson equation in Lehmann representation
[\onlinecite{gw1}]. $V_{H}$ is the Hartree potential of the QM part,
$U^{QM}$ is the electron-ion potential of the QM part,
while $U^{QM/MM}$ is the potential felt by the electrons due to the point
charges of the MM part. Finally, $\varepsilon _{j}^{QP}$ are the
quasi-particle eigenvalues. 
Eq.\ (\ref{dyson}) has the same form as the KS equation
[\onlinecite{dft2}] in the presence of an external electric field,
where the exchange-correlation potential $V_{xc}(\mathbf{r}$) is
replaced by the self energy $\Sigma (\mathbf{{r},{r^{\prime
}}},\varepsilon _{j}^{QP})$ which acts as a non-local,
energy-dependent potential.
Therefore, the eigenvalue problem described above can be solved
perturbatively considering the KS equation as an unperturbed Hamiltonian and
$\Sigma -V_{xc}$ as a perturbative term. The quasi-particle eigenvalues are
obtained in first order approximation: 

\begin{equation}
\varepsilon^{QP}_{j} = \varepsilon^{KS}_{j} + \frac{\langle
\phi_{j}^{KS}|\Sigma(\varepsilon_{j}^{KS}) - V_{xc}|\phi_{j}^{KS}\rangle} {%
1-\langle \phi_{j}^{KS}|\frac{d \Sigma(\varepsilon^{KS}_{j})}{d \omega}%
|\phi_{j}^{KS}\rangle}.  \label{first_perturb}
\end{equation}

All the Coulomb interactions, and hence also the one induced by the
classical region, are included in the KS eigenvalues $\varepsilon^{KS}_{j}$,
and eigenvectors $|\phi_{j}^{KS}\rangle$. In this GW/MM scheme we neglect
the contribution of the classical atoms to the $\Sigma$ operator in the same
way as it is neglected for $V_{xc}$ in the DFT/MM scheme.

As a bonus from GW calculations we obtain also the four-point
independent quasi-particle polarizability required
 to solve the BSE
and which reads, in transition space [\onlinecite{bse}]:

\begin{equation}
P^{0}_{(n1,n2)(n3,n4)}(\omega) = \frac{f_{n2}-f_{n1}}{\varepsilon^{QP}_{n2}-%
\varepsilon^{QP}_{n1}-\omega} \delta_{n1,n3} \delta_{n2,n4}
\label{polarizability}
\end{equation}
where $f_{n}$ is the occupation numbers of level $n$: the quantum
plus classical external potential is not explicitly present in the
BSE, but indirectly determines all the ingredients.

\begin{figure}[h]
\includegraphics[width=8.5cm,angle=0]{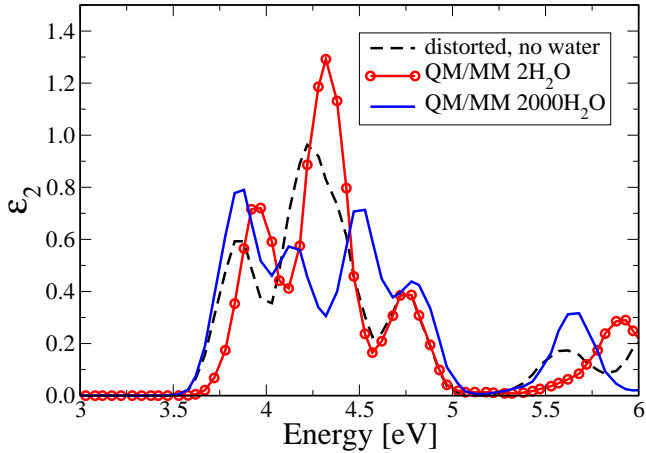}
\caption{(color online) DFT-IPA optical  spectra of indole in 2000 water
molecules (blue solid line), in 2 water molecules (red circles) and without
water (black dashed line).}
\label{2w_qmmm}
\end{figure}

We performed a 20 ps hybrid QM/MM Car-Parrinello simulation of an
indole molecule (QM part) [\onlinecite{detailsDynamics}] surrounded
by 2000 water molecules treated classically (MM part), with the
Amber force field [\onlinecite{detailsMM}]. Such a large number of water
molecules was necessary to correctly reproduce the physical properties of a
disordered system such as liquid water at room temperature (300 K). We also
verified that such a large number of water molecules used for the dynamics
was also necessary to calculate the optical absorption spectrum. This became
clear already at the DFT-independent particle approach (DFT-IPA), as shown in
Fig.\ref{2w_qmmm}: we calculated the absorption spectrum for indole with no
water (with indole in the same "distorted" geometrical configuration as if in water), for indole with 2 water molecules, and for indole with 2000 water molecules. The three spectra are all different. This is a strong indication that also for the absorption spectrum it is necessary to include many water molecules in the theoretical simulation. This is due to the long range electrostatic potential of water which acts on indole.

Next, we tested our assumption that the solvent can be treated
classically in the calculation of absorption spectra by performing
TDDFT calculations [\onlinecite{detailsTDDFT}] for a system where
two water molecules were treated at quantum level, and the remaining
1998 classically.
The position of absorption peak is the same (within 0.02 eV) as in
the case where all waters were treated within MM. Our result
supports the use of this approach for solutes in water
[\onlinecite{tddftmm2}].

For ten snapshots of the QM/MM dynamics (one every two ps) we
computed the optical spectra at the independent particle level
(DFT-IPA) and within TDDFT. The final spectrum was obtained by an
average over the snapshots. The convergence of the spectrum was
reached already after 6 snapshots, hence the subsequent GW and BSE
calculations have been performed on only 6 snapshots.

The calculated DFT and GW HOMO-LUMO gaps [\onlinecite {detailsGW}],
averaged over the QM/MM configurations, are 3.8 eV (with standard deviation $\sigma=\pm$ 0.1 eV)
and 7.2($\sigma=\pm$ 0.2) eV, respectively. The GW
corrections to the gap turned out to be practically constant in all
the snapshots considered (3.4 $\pm$ 0.1 eV).
This fact, already found for liquid water
[\onlinecite{viviana}], confirms that one can strongly reduce the
computational effort, by performing a GW calculations for just one
snapshot.

\begin{figure}[h]
\begin{center}
\begin{tabular}{c}
\includegraphics[width=8.5cm,angle=0]{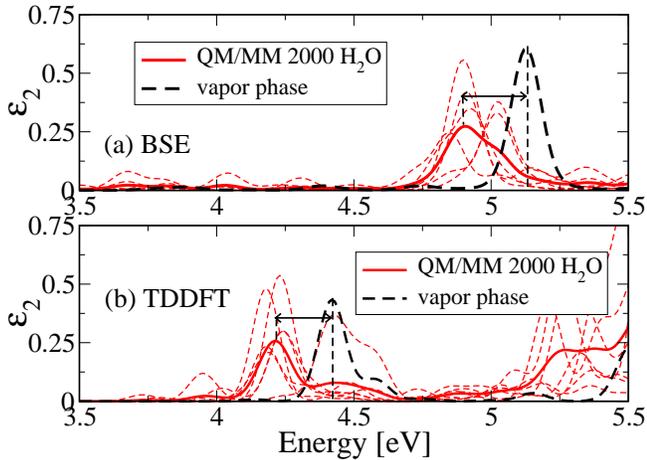} \\
\end{tabular}%
\end{center}
\caption{(color online) BSE (a) and TDDFT (b) spectra of indole in water.
The tiny red dashed lines are  the
spectrum of each snapshot. The red solid line is obtained by an average
over these spectra. The black dashed line is for indole in vapor phase.}
\label{solventshift}
\end{figure}
\begin{figure}[tbp]
\includegraphics[width=8.5cm,angle=0]{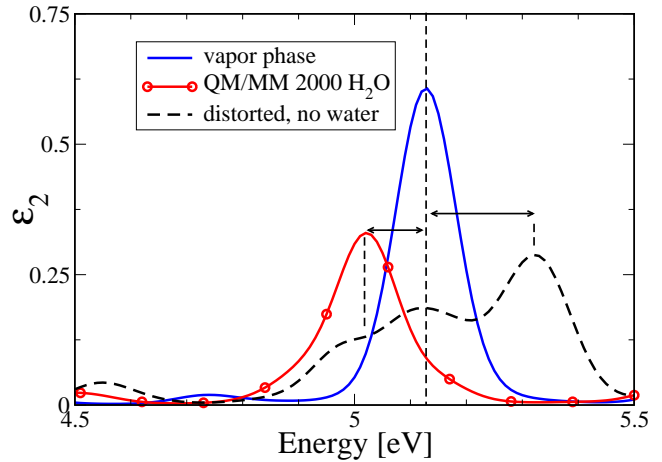}
\caption{(color online) BSE optical spectra.
 Solid blue line:
Indole
in vapor phase. Black dashed line: indole without water
molecules, with distorted geometry   taken from a snapshot corresponding to
13.08
ps of the dynamics. Circles: indole in water,  spectrum calculated for
the same snapshot.}
\label{fig-bse-geow}
\end{figure}

We finally calculate the low energy range of the optical spectrum of
indole, by GW-BSE and TDDFT, always as result of an average over the
QM/MM snapshots. In Fig.\ref{solventshift} we report our results
together with the calculated absorption spectrum in gas phase. We
notice that, in both approaches the most intense peak ($^{1}L_{a}$
in the experiment) is red-shifted on passing from gas phase to the
water solution. This agrees with experiments
[\onlinecite{exp_indole1}]. Same conclusion was obtained by previous
theoretical calculations of indole in water based on CASSCF method
and CASPT2 [\onlinecite{exp_indole2}]. In these approaches, the
solvent was simulated by a continuum model with a cavity containing
the indole molecule. The value we calculate for such a redshift is
$\sim$ 0.2 eV, in fair agreement with experiment (0.18 eV)
[\onlinecite{exp_indole1}]. On the contrary the CASSCF and CASPT2
prediction for the solvent shift is about 0.06 eV only. Such an
underestimation may depend on the geometrical distortion of indole
molecule caused by temperature effects due to the solvent and by an
explicit H-bonding between water molecules, which were not
considered explicitly therein. To quantify the effect of the
geometry distortion on such shift, we performed calculations of
indole switching on and off the water field in order to separate the
geometry effect from the electrostatic ones. The results are
presented for a single snapshot in GW+BSE (Fig. \ref{fig-bse-geow})
and are obtained by performing additional calculations for the same
QM/MM configuration without the water field. The corresponding
solvent-shift goes from -0.1 eV with water field to +0.2 eV (hence,
a blue-shift) without water field. This emphasises the importance of
taking into account explicitly  the electrostatic interaction with
the solvent, since the geometry distortion alone would give, at
least for this snapshot, a wrong sign.

In addition, TDDFT underestimates the energy of the $^1L_a$ peak both in  gas phase and in solution
 by $\sim $0.4 eV, and BSE-GW overestimates them by
  $\sim $0.3 eV.
As expected
[\onlinecite{exp_indole3}], CASSCF is much worse, it overestimates
by $\sim $1 eV or more, whilst CASPT2 is more accurate ($\sim $0.13
eV or less).

In conclusion, we have included many-body perturbative techniques in
a QM/MM scheme. We have applied it, together with a TDDFT/MM
approach, to study the optical properties of indole in water
solution. Both methods reproduce quantitatively the redshift induced
by the solvent. Hence, the GW-BSE method could be applied to
biomolecules in aqueous solution (i.e. in laboratory-realizable
conditions) in cases where the TDLDA/GGA approach does not work
[\onlinecite{failure1,failure2}]. Our GW-BSE calculations further
show that the solvent shift is a consequence of the combination of
two effects: the geometrical distortion of indole molecule in the
solvent and the electrostatic interaction with the water molecules
electric dipoles. Both effects, and their sum, depend on the
particular configuration of the system; this emphasizes the need of
more than one snapshot (several, indeed) for carrying out accurate
optical calculations.

This work opens the way to further applications in other bio-relevant
molecules, such as proteins and cell membranes, for which the evaluation of
the optical shift enables to understand the nature of their environment.

This work was supported by the EU through the Nanoquanta NOE
(NMP4-CT-2004-500198). Computer resources from INFM ``Progetto Calcolo
Parallelo'' at CINECA are gratefully acknowledged. We also thank L.
Guidoni for interesting discussions.


\begin{thebibliography}{99}
\bibitem{detailsDynamics} The QM/MM code combines
the CPMD3.11.1 [\onlinecite{cpmd}] and Gromos [\onlinecite{gromos}]
codes. The time step chosen for the simulations was about 0.1 fs.
Since the indole's N--H group forms H-bonds with water, we used an
energy cutoff of 70 Ry with the BLYP recipe [\onlinecite{blyp}] for
the exchange-correlation functional, which have widely been used for
biophysical applications. Norm-conserving pseudopotentials of
Troullier-Martins type [\onlinecite{TM}] corrected for a better
description of Van der Walls interactions [\onlinecite{VdW}] have
been used. A Nose-Hoover thermostat [\onlinecite{nose,nose1}] is
applied throughout all simulations to keep the temperature constant.
\bibitem{detailsMM} We used the Parm99 version of Amber force field with
the TIP3P model [\onlinecite{tip3p}] for water molecules.
\bibitem{detailsTDDFT} TDDFT calculations are obtained within
the Tamm-Dancoff approximation [\onlinecite{TD1,TD2}] as
implemented in CPMD 3.11.1 package, using the PBE [\onlinecite{PBE}]
exchange-correlation functional. The same corrected Troullier-Martins
pseudopotentials as for the dynamics have been used.
\bibitem{detailsGW}
GW calculations have been done by using 12077 plane-waves and 500
electronic bands.
Following [\onlinecite{onida}],
we used a cutoff in the real space for the self energy to prevent
that periodic images of the quantum part interact with each other.
The screen function is calculated within a plasmon pole
approximation.
\end{thebibliography}
\end{document}